\documentclass{elsart}

\usepackage{graphicx}
 \usepackage{epsfig}

\usepackage{amssymb}

\begin{document}

\begin{frontmatter}



\title{
Decentralised control of \\material or traffic flows in networks \\
using phase-synchronisation }

\author[1]{Stefan L\"ammer},
\author[2]{Hiroshi Kori},
\author[1] {Karsten Peters}, and
\author[1,3]{Dirk Helbing}

\address[1]{
    Technische Universit\"{a}t Dresden, \\
    Andreas-Schubert-Str. 23, D-01062 Dresden, Germany}

\address[2]{
    Fritz-Haber-Institut der Max-Planck-Gesellschaft,\\
    Faradayweg 4-6, D-14195 Berlin, Germany}

\address[3]{
    Collegium Budapest -- Institute for Advanced Study, \\
    Szenth\'aroms\'ag utca 2, H-1014 Budapest, Hungary}

\begin{abstract}
We present a  self-organising, decentralised control method for
material flows in networks. The concept applies to networks where
time sharing mechanisms between conflicting flows in nodes are
required and where a coordination of these local switches on a
system-wide level can improve the performance. We show that, under
certain assumptions, the control of nodes can be mapped to a network
of phase-oscillators.
\\
By synchronising these oscillators, the desired global coordination
is achieved. We illustrate the method in the example of traffic
signal control for road networks. The proposed concept is flexible,
adaptive, robust and decentralised. It can be transferred to other
queuing networks such as production systems. Our control approach
makes use of simple synchronisation principles found in various
biological systems in order to obtain collective behaviour from
local interactions.
\end{abstract}


\end{frontmatter}
\emph{If you want to cite this report, please use the following
reference instead:} \\
S. L\"ammer, H. Kori, K. Peters, and D. Helbing (2006)
\\Decentralised control of material or traffic flows in networks\\
using phase-synchronisation, \emph{Physica A} {\bf 363}(1) pp. 39-47
 \pagebreak

\section{Introduction}
\label{sec:into}

Efficient control of multiple material flows on a system wide level
is one of the most challenging problems in complex socio-technical
transportation systems.
\\
For the majority of all material flow systems, a parallel service of
several intersecting flows or conflicting tasks is impossible,
unsafe or inefficient. Alternating exclusion of competing tasks is
frequently observed at crossroads in road traffic
\cite{Papa91,HelbingLaemmer05,aut-dia03,Gershenson04,aut-rob91,SekiyamaOhashi05,aut-hen90},
in the organisation of production processes
\cite{Radons,Pe04,RemArmbruster02,Chen01} or in communication
networks \cite{comnet1,comnet2}. Instead of parallel processing of
different flows, a sequential processing must be organised in an
optimal manner. Whereas the setup of optimal schedules for single
nodes is usually done by traditional optimisation techniques, the
control problem on a network-wide level becomes often practically
unsolvable by these methods, especially in larger networks.
\\
Moreover, most of the material flow networks are subject to
continuous demand variations and unforeseen failures. Besides
adaptivity and optimality, robustness and flexibility are important
requirements for control concepts.
\\
Can we learn from the stable, smooth, and efficient flow of
nutrients and other chemical substances in circulatory systems of
biological organisms? Synchronised dynamics of a population of cells
often plays an important role for it
\cite{pikovsky01,winfree80,Cam01}. For example, our heart functions
as a pump through the appropriate synchronised dynamics of a
population of cardiac cells. This synchronisation is realised
through appropriate designs of cardiac cells and their network
architecture of local interactions. Another interesting example is
found in amoeboid organisms \cite{nakagaki00,tero05}, where the
rhythmic contraction pattern produces streaming of protoplasm.
Synchronisation phenomena have been intensively studied for these
biological systems during the last decades by means of mathematical
models, in particular coupled phase-oscillator models
\cite{pikovsky01,blek88,ermentrout91,kuramoto84,Strogatz01}. The
great advantage of phase-oscillator models are their tractability
and universality.

In the present paper, we take advantage of this tractability and
propose a decentralised control principle for material flow networks
with transportation delays and setup-times, based on the
phase-synchronisation of  oscillatory services at the network nodes.

\subsection{Decentralised control using phase-synchronisation}
\label{sec:concept}

Let us consider a material transport network, which is a directed
graph with of a set of nodes and links. Material can move between
the nodes with a finite velocity. Thus, any element experiences a
delay $t_{ij}$ between its departure at one node $i$ and its arrival
at the next node $j$. Whereas a distinct subset of nodes may act as
a source or sink of moving material, we shall concentrate on those
nodes where the flow of material is conserved, i.e.
\begin{equation}
    \label{eq:conservation1}
    \sum q^\mathrm{in} = \sum q^\mathrm{out}.
\end{equation}
Here  $\sum q^\mathrm{in}$ and $\sum q^\mathrm{out}$ denote the
average rate of incoming and outgoing material, respectively. Each
node has to organise the routing of materials arriving through
incoming links towards its outgoing links. All allowed connections
between incoming and outgoing links can be described through
discrete states of the respective node. As long as such a state is
`active', material can flow from a subset of incoming links through
the node and leave through outgoing links. All other flow relations
are blocked. Usually the switching between different discrete states
needs a certain time interval $\tau$, called switching- or
setup-time. Depending on the flow rates, the duration of these
discrete states may vary. Since we assume a cyclic service sequence,
we can assign a periodic motion to every switching node. Thus, a
node can be modelled as a hybrid system consisting of a
phase-oscillator and a piecewise constant function $M$ that maps the
continuous phase-angle $\varphi(t)$ to the discrete service state
$s(t)$, e.g. $M:\varphi(t)\rightarrow s(t)$.
\\
The switched service of different flows leads to convoy formation
processes. This implies highly correlated arrivals at subsequent
nodes, which requires to optimise $M$ with respect to a minimal
delay of the material. Whereas the map $M$ can be optimised
according to the actual local demand, the phase-angle $\varphi$ is
coupled to the oscillatory system of the neighbouring nodes. Thus,
with a suitable synchronisation mechanism we can achieve a
coordination of the switching states on a network-wide level. In
consequence, we suggest an adaptive decentralised control concept
consisting of two parts:
\begin{enumerate}
\item[(a)]
Phase-synchronisation of all oscillators in the network, based on
local coupling between immediate neighbours.
\item[(b)]
Mapping of phase-angles to the switching states based on local
optimisation.
\end{enumerate}
For the sake of concreteness, we apply our  method to the control of
traffic lights at intersections of road networks. The rest of this
paper is organised as follows: At first, in Sec.~\ref{sec:model} we
propose a hybrid phase-oscillator model for the intersections of a
road network. In Sec.~\ref{sec:synchro} we discuss the
synchronisation of the oscillators by a suitable local coupling
mechanism for arbitrary network architectures. Finally, in
Sec.~\ref{sec:traffic} we show how an optimal switching of green and
red traffic light phases with respect to minimum delay times can be
reached, based on local adjustment and optimisation.

\section{Hybrid model for traffic light control}
\label{sec:model}

In the following, we introduce a hybrid oscillator model for a
single traffic light controlled intersection and derive an upper
bound for the allowed oscillator frequency.

\subsection{Model of an intersection}

\begin{figure}[t]
    \begin{center}
        \begin{picture}(200,200)(0,0)
        \includegraphics[width=200pt]{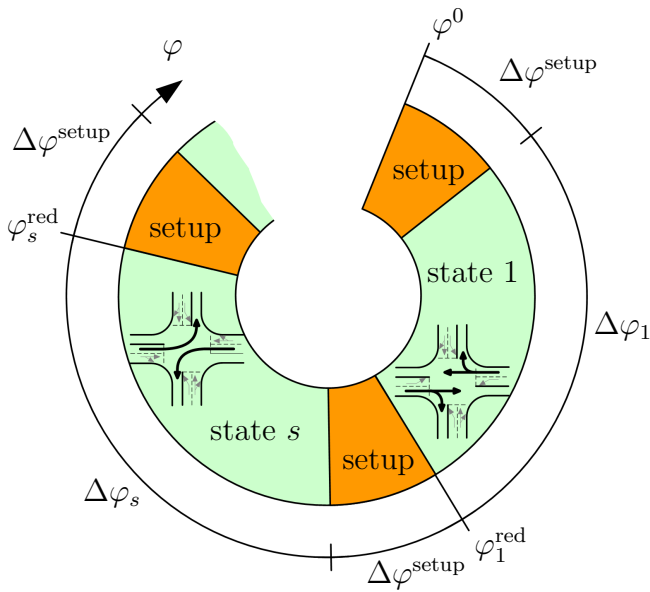}
        \end{picture}
        \begin{picture}(0, 200)(200, 0)
            \put(122,149){setup}
            \put(103,39){setup}
            \put(30,127){setup}
            \put(135,110){state 1}
            \put(53,50){state $s$}
            \put(137,204){$\varphi^0$}
            \put(162,186){$\Delta \varphi^\mathrm{setup}$}
            \put(198,90){$\Delta \varphi_1$}
            \put(153,7){$\varphi_1^\mathrm{red}$}
            \put(112,-4){$\Delta \varphi^\mathrm{setup}$}
            \put(5,27){$\Delta \varphi_s$}
            \put(-24,128){$\varphi_s^\mathrm{red}$}
            \put(-22,160){$\Delta \varphi^\mathrm{setup}$}
            \put(35,196){$\varphi$}
        \end{picture}
    \end{center}
    \caption{
A single intersection adjusts the mapping of the phase-angle
$\varphi$ to the switching states $s$ locally. Within a complete
cycle, each state $s$ is sequentially activated for a period $\Delta
\varphi_s$, during which the corresponding non-conflicting traffic
lights are set to green. While switching from one state to another,
all traffic lights are set to red for a period of $\Delta
\varphi^\mathrm{setup}$. The phase-angle, at which a new cycle
starts, is denoted by $\varphi^0$.
    }
    \label{fig:cycle}
    \vspace{10mm}
\end{figure}

An intersection in road traffic  is given by a set of traffic lights
$l \in \mathcal{L}$, each one controlling the vehicular flow of a
single or several lanes. If two driving paths intersect, we call the
related green lights conflicting. We require that conflicting
traffic lights are not set to green at the same time. In order to
increase the throughput of an intersection, non-conflicting traffic
lights can be switched collectively. Such a collective switching of
a subset $\mathcal{L}_s \subset \mathcal{L}$ of traffic ligths
corresponds to one discrete state $s\in \{1,2,\ldots S\}$ of the
intersection. An explanatory sketch of this model is shown in Fig.
\ref{fig:cycle}.
\\
While the phase-angle $\varphi$ of the oscillator modelling the
intersection progresses from $0$ to $2\pi$ at the rate $\omega$, all
states $s$ are sequentially activated for some period $\Delta
\varphi_s$. Thus, all traffic lights in the subset $\mathcal L_s$
are set to green for this interval $\Delta \varphi_s$ whereas the
remaining lights $l \in \mathcal{L} \setminus \mathcal{L}_s$ are
turned to red. Moreover, for switching from one state to another,
all traffic lights must be set to red for a time-period of $\Delta
\varphi^\mathrm{setup}(\omega) = \tau \omega$ due to security
reasons \cite{Papa91}. These intervals correspond to a switching- or
setup-time $\tau$  (see Fig.~\ref{fig:cycle}). In conclusion, we
obtain
\begin{equation}
    \label{eq:cycle}
       \sum_{s=1}^S
    \left(
        \Delta \varphi^\mathrm{setup} + \Delta \varphi_s
    \right)
    =    2\pi.
\end{equation}

\subsection{Maximum oscillator frequency}
\label{sec:omegaMax}

To fulfil the balance condition Eq.~(\ref{eq:conservation1}) we have
to guarantee that the number of vehicles arriving during a cycle  at
the intersection is equal to the number of vehicles that can pass
the intersection and depart.
\\
The instantaneous vehicular flow $q_l(\varphi)$ is given by the
number of vehicles $dn_l$ that `potentially' arrive at traffic light
$l$ during $(\varphi, \varphi + d\varphi)$, which is equal to the
time-shifted flow measured at a cross-section sufficiently far
upstream the traffic light.\\
In the following we do not require any model for the dynamics of
$dq_l$ thus the control concept is independent of the underlying
traffic model. Since the traffic situation usually changes slowly
compared to the period of a full switching cycle we assume the
average flows $\bar q_l = 1/2\pi \int_0^{2\pi} q_l(\varphi)
d\varphi$ to be constant. The departure rate during green light is
$q_l^\mathrm{max}$ if there are waiting vehicles, and $q_l(\varphi)$
otherwise. If the traffic light shows red, the departure rate is
zero. As a consequence, we obtain a lower bound for the green time
periods $\Delta \varphi_s$.
\begin{equation}
    \label{eq:QueueStability}
    \Delta \varphi_s
    \;\geq\;
    2 \pi
    \; \max_{l \in \mathcal{L}_s}
    \; \bar q_l / q_l^\mathrm{max}.
\end{equation}
Here $ \bar q_l / q_l^\mathrm{max}$ is the utilisation of the lanes,
which are served during state $s$. For the signal control to
function well, we require $\bar q_l / q_l^\mathrm{max} < 1$. If $q_l
= q_l^\mathrm{max}$ for some $l \in \mathcal{L}_s$, we have $ \Delta
\varphi_s=2\pi$, which implies that state $s$ must be set to green
all the time in order to allow for the departure of all vehicles
arriving from lane $l$. In this case, our synchronisation strategy
would fail to switch to other states, while a purposeful operation
of an intersection requires a switching through all $S$ states in
one cycle.
\\
The load of the whole intersection is determined by
\begin{equation}
    \label{eq:u}
    u =
    \sum_{s=1}^S
    \; \max_{l \in \mathcal{L}_s}
    \; \bar q_l / q_l^\mathrm{max}.
\end{equation}
Additionally, the conditions imposed by
Eq.~(\ref{eq:QueueStability}) and Eq.~(\ref{eq:cycle}) must be
satisfied. This is guaranteed only if
\begin{equation}
\label{eq:omegaMax}
 \omega\le
    \frac{2 \pi}{S \tau} \left( 1-
    \sum_{s=1}^S \max_{l \in \mathcal{L}_s} \bar q_l / q_l^\mathrm{max}\right)
    \;\;=:\;\;
    \omega^\mathrm{max}=
    \frac{2 \pi}{S \tau}
    \big( 1 - u \big)
    \;.
   \end{equation}
In other words, there exists an upper bound $\omega^\mathrm{max}$
for the switching frequency, determined by the amount of time which
is spent on setups during a cycle while the balance condition
Eq.~(\ref{eq:conservation1}) is fulfilled. For an increasing load
$u$, the overall time for setups must be reduced relative to the
duration of a cycle. This suggests to decrease $\omega^\mathrm{max}$
for increasing loads $u$. If $u = 1$, the maximum frequency becomes
$\omega^\mathrm{max} = 0$. In this case, the intersection is
blocked, since there exists no switching cycle which allows all
arriving vehicles to depart. This is the typical behaviour of all
queuing systems, if the inflow reaches the maximum
capacity\footnote{In our case, the capacity of an intersection is
determined by the maximum departure rates $q_l^\mathrm{max}$, the
switching times $\tau$ and the partition $\{\mathcal{L}_s\}$.}.
Small values of the load $u$ allow for a higher maximum frequency
$\omega^\mathrm{max}$ and perhaps also a higher cycle frequency
$\omega \le \omega^\mathrm{max}$, which is in favour of smaller
delays of vehicles. This is obvious, as the maximum delay time of a
vehicle is given by the duration of a full cycle.

\section{Global coordination of intersections by synchronisation}
\label{sec:synchro}

The objective of our decentralised control method is the network
wide coordination of the individual switching sequences based on a
local coupling between the intersections in the road network. By
modelling each intersection $i$ as an oscillator, characterised by
its phase-angle $\varphi_i$ and its effective frequency $\omega_i =
\dot \varphi$, coordination is achieved by synchronising the
oscillator network. Hereby, for providing a common time-scale and
allowing the intersections to trigger the switching cycles right at
the best time (see Sec.~\ref{sec:traffic}), we require a
phase-locked state where the phase-difference between neighbouring
oscillators is fixed \cite{pikovsky01}.
\\
Therefore, we suggest a coupling between any oscillator
$i=1,2,\ldots N$ and its nearest neighbours $j \in \mathcal{N}_i$
with adjustments of phases and frequencies on two different
timescales.
\\
At first we consider the adaption of the phase-angle $\varphi_i$:
\begin{equation}
    \label{eq:PhaseSync}
    \dot{\varphi}_i
    =
    \min\Big\{\;
    \omega_i^\mathrm{max}
    ,\;\;
    \Omega_i(t) +
    \frac{1}{T_\varphi}
    \sum_{j \in \mathcal{N}_i}
    \sin\big( \varphi_j(t) - \varphi_i(t) \big)
    \;\Big\}
    =:
    \omega_i(t)
\end{equation}
where $\Omega_i$ is the inherent frequency. As long as $\omega_i <
\omega_i^\mathrm{max}$, $\varphi_i$ tries to adjust to the
neighbouring phase-angles $\varphi_j$. The constant $T_\varphi$
corresponds to the typical time-scale for this adaption.

Beyond the phase-synchronising interaction given by $\sin(
\varphi_j(t) - \varphi_i(t))$, a second decentralised coupling can
be used to increase the inherent frequencies to approach the
possible maximum within a slow time-scale:
\begin{equation}
    \label{eq:FrequSync}
    \dot{\Omega}_i
    =
    \frac{1}{T_\Omega}
    \Big(
        \;
        \min_{j \in \mathcal{N}_i} \big\{ \omega_j(t) \big\}
        + \Delta \Omega
        \;
        - \Omega_i(t)
        \;
    \Big)
    \;.
\end{equation}
Here the constant parameter $\Delta \Omega > 0$  provides a linear
drift towards higher frequencies.
\subsection{Synchronisation dynamics and network size}
\label{sec:syncDynamics}

\begin{figure}[t]
    \begin{center}
        \includegraphics[width=\textwidth]{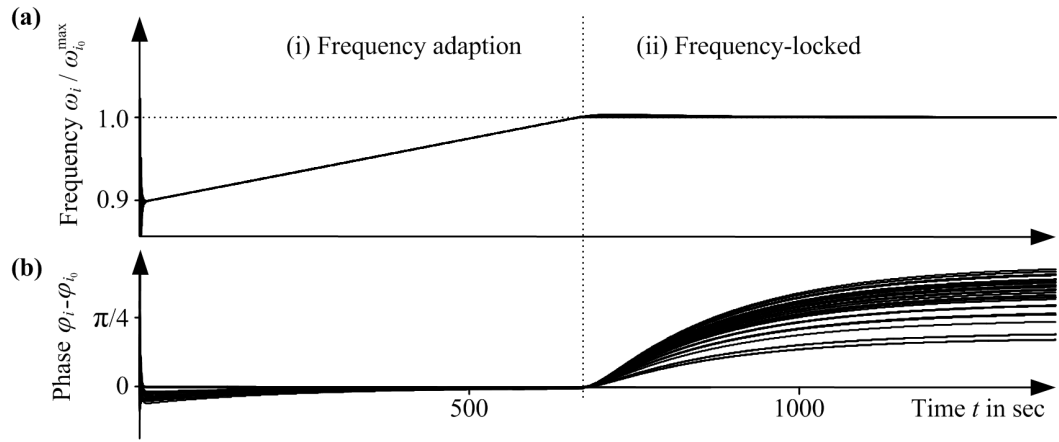}
    \end{center}
    \caption{
Simulation results for a regular lattice road network, where the
$5\times 5$ intersections are defined as oscillators with (a) a
frequency $\omega_i$ and (b) a phase-angle $\varphi_i$. Based on a
local coupling between immediate neighbours, all oscillators in the
network synchronise globally by adjusting their phase-angles
mutually. Left: Starting with a random initial condition, the system
quickly approaches state (i) with a steadily growing common
frequency and vanishing phase-differences. Right: As soon as the
maximum possible common frequency is found (indicated by the
horizontal dotted line), the system enters state (ii) with a locked
common frequency and phase-differences exponentially converging
towards constant values. ($T_\varphi$ = 300s, $T_\Omega$ = 60s,
$\omega_{i_0}^\mathrm{max} = 2\pi/$60s,
$\Delta\Omega/\omega_{i_0}^\mathrm{max} = 10^{-3}$, and $N=25$)
    }
    \label{fig:synchro-states}
    \vspace{10mm}
\end{figure}

If the coupling mechanisms outlined above are applied to a network
of oscillators, two different adaptive behaviours of the system can
be distinguished. Either, the system can evolve freely and increase
the common frequency, i.e. a slow {\it frequency adaption} is
possible. Otherwise, the possible dynamics of the coupled system is
restricted by an intersection $i_0$, whose maximum frequency is
reached. This requires the remaining oscillators to be {\it
frequency-locked} to $\omega_{i_0}^\mathrm{max}$.
Fig.~\ref{fig:synchro-states} shows an example of both dynamical
regimes which are discussed in the following.

(i) {\em Frequency adaption}.\\
As long as $\omega_i < \omega_i^\mathrm{max}$, a synchronised
solution is obtained:
\begin{equation}
     \label{eq:state1}
     \varphi_i(t) = \varphi_j(t), \quad
     \omega_i(t) = \Omega_i(t), \quad \mathrm{and}\quad
     \dot \Omega_i = \frac{\Delta \Omega}{T_\omega}
    \quad \forall i\;.
\end{equation}
All oscillators have exactly the same phase-angle, while the
frequencies $\omega_i$ and $\Omega_i$ of all oscillators $i$
increase linearly with time. As soon as any oscillator reaches its
maximum frequency, however, Eq.~(\ref{eq:PhaseSync}) is dominated by
$\omega_i^\mathrm{max}$ in the minimum function and $\omega_{i}$
will not increase anymore. Instead, the other state appears.

(ii) {\em Frequency-locked state}.\\
In this state, the effective frequency $\omega_{i_0}$ of the
oscillator $i_0$ with the global minimum of all maximum frequencies
is locked to $\omega_{i_0}^\mathrm{max} = \min_{i'}
\omega_{i'}^\mathrm{max}$. In order to achieve synchronisation, the
effective frequencies $\omega_i$ of all oscillators must become
$\omega_{i_0}^\mathrm{max}$. Under suitable conditions derived
below, the frequency-locked state has the solution
\begin{equation}
     \label{eq:state2}
     \omega_i
     = \omega_{i_0}^\mathrm{max}
     = \min_{i'} \omega_{i'}^\mathrm{max}
     \quad \mathrm{and}\quad
     \Omega_i = \omega_{i_0}^\mathrm{max} + \Delta \Omega
     \quad \forall i\;.
\end{equation}
The drift parameter $\Delta \Omega$, which was in the previous state
(i) responsible for increasing frequencies, must now be compensated
by the phase-differences. Using Eqs.~(\ref{eq:PhaseSync}) and
(\ref{eq:state2}) leads to:
\begin{equation}
     \label{eq:existence1}
     \sum_{j \in \mathcal{N}_i}
     \sin\big( \varphi_j(t) - \varphi_i(t) \big)
     \;    =    \;
     - T_\varphi \, \Delta\Omega
     \quad    \forall i \neq i_0\;.
\end{equation}
Furthermore
\begin{equation}
     \label{eq:existence2}
     \sum_{i=1}^N
     \sum_{j \in \mathcal{N}_i}
     \sin\big( \varphi_j(t) - \varphi_i(t) \big)
     \;    =    \;    0\;,
\end{equation}
which is due to the anti-symmetry of the sin-function. Therefore,
the phase-angles of the oscillator $i_0$ and its next neighbours $j
\in \mathcal{N}_{i_0}$ must satisfy
\begin{equation}
    \label{eq:existence3}
    \sum_{j \in \mathcal{N}_{i_0}}
    \sin\big( \varphi_j(t) - \varphi_{i_0}(t) \big)
    \;    =    \;    (N - 1) \, T_\varphi \, \Delta\Omega\;.
\end{equation}
Considering an arbitrary network of $N$ oscillators, there always
exists a solution for the immediate neighbours of $i_0$ if
\begin{equation}
     \label{eq:existenceCondition}
     (N - 1) \, T_\varphi \, \Delta\Omega\;\leq\;1\;.
\end{equation}
A similar condition can be derived for the second-next neighbours
and so forth, but for these, the factor in place of $(N-1)$ will
become smaller and smaller. Therefore,
Eq.~(\ref{eq:existenceCondition}) gives a sufficient condition for
the existence of a phase-locked solution in the
frequency-locked state (ii).\\
Equation~(\ref{eq:existenceCondition}) exposes an compromise between
large network sizes and a fast adapting behaviour. Increasing the
network size $N$ requires a smaller adaption rate $\Delta \Omega$ to
ensure the synchronisation and vice versa.
\\
From a given initial condition, the system always converges to the
phase-locked state (ii), see Fig.~ \ref{fig:synchro-states}. It
might temporarily enter the frequency adaptation state (i), where
the common frequency grows in time to approach the frequency of the
slowest oscillator. The transition between states (i) and (ii),
which occurs when one of the oscillators reaches its maximum
frequency, is smooth. There are no jump-like disturbances in
effective frequencies and phase-angles. Once the system converges to
state (ii), the synchronised frequency tightly follows variations of
the global minimum of $\omega_i^\mathrm{max}$ frequencies. The
phase-differences between neighbouring oscillators are almost
constant. Thus, the phase-synchronisation establishes a
load-adaptive common framework among the locally coupled
oscillators, which can be used to coordinate the individual
switching sequences on a network-wide level.

\section{Control of periodic traffic flows at a single intersection}
\label{sec:traffic}

The time needed by a vehicle to traverse a series of roads
controlled by traffic lights is given by the travel times between
the intersections and the delay times at red lights. To minimise the
overall travel time, or, to increase the throughput of the
controlled network, we shall minimise the delay times at every
intersection. According to the synchronisation concept outlined in
Sec.~\ref{sec:synchro}, each intersection $i$ is provided with a
phase-angle $\varphi_i$, phase-locked to its neighbours and
synchronised to a common frequency $\omega$. Due to frequency
synchronisation in steady state, the vehicular flows are basically
periodic functions of the phase-angle $\varphi_i$. Each intersection
$i$ may now independently adjust its switching sequence
$M:\varphi(t)\rightarrow s(t)$  based on the local arrivals of the
vehicles. Thus, for clarity, we neglect the index $i$ in this
section and express time $t$ in terms of $\varphi$.
\\
We  shall answer the question, where in the cycle the state $s$
should start, to minimise the total time that all vehicles may need
to pass the intersection. Therefore, by $\varphi_s^\mathrm{red}$ we
denote the point where the subset $\mathcal{L}_s$ of traffic lights
switches to red after a green time period $\Delta \varphi_s$.
\\
The delay of a single vehicle is given by the period between its
arrival $\varphi^\mathrm{arr}$ and its departure
$\varphi^\mathrm{dep}$. A vehicle may arrive at the point
$\varphi^\mathrm{arr} = \varphi_s^\mathrm{red} + 2\pi p$ with
$0<p\le 1$. Since the vehicles are not rearranged during the queuing
process, it is expected to depart at $\varphi^\mathrm{dep} = 2\pi +
\varphi_s^\mathrm{red} - (1 - p) \Delta \varphi_s$. The resulting
delay of the particular vehicle is therefore
\begin{equation}
    \label{eq:singleDelay}
    \varphi^\mathrm{dep} - \varphi^\mathrm{arr}
    =
    (2\pi-\Delta \varphi_s)(1-p)\;.
\end{equation}
Integrating the delays of all vehicular flows $q_s(\varphi) =
\sum_{l \in \mathcal{L}_s} q_l(\varphi)$ served in a state $s$ over
a complete cycle gives the delays $D_s$ imposed by the subset
$\mathcal{L}_s$ of traffic lights as a function of
$\varphi_s^\mathrm{red}$ and $\Delta \varphi_s$:
\begin{equation}
    \label{eq:TotalDelayState}
    D_s ( \varphi_s^\mathrm{red}, \Delta \varphi_s )
    \;=\;
    \int_0^1
    \;(2\pi-\Delta \varphi_s)(1-p)
    \;q_s
    ( \varphi_s^\mathrm{red} + 2 \pi p)
    \;2\pi dp.
\end{equation}
Now we have to minimise the overall delay
\begin{equation}
D = \sum_{s=1}^S
D_s\left(\varphi_s^\mathrm{red}, \Delta\varphi_s\right)
\label{eq:delayCost}
\end{equation}
produced by an intersection during a cycle.
\\
To identify the optimum value of $\varphi_s^\mathrm{red}$ and
$\Delta \varphi_s$ for all  $\mathcal{L}_s \subset \mathcal{L}$ we
consider the following two assumptions:
\\
(i) The sequence of states is given, e.g. by the order of their
indices $s=1 \ldots S$.
\\
(ii) The periods $\Delta \varphi_s$ are fixed and fulfil the
conditions Eqs.~(\ref{eq:cycle}) and (\ref{eq:QueueStability}).
\\
Then there remains only one degree of freedom, which can be
expressed by the phase-angle $\varphi^0$ that triggers the switching
to state $s=1$ (see Fig.~\ref{fig:cycle}). Thus, the end-point
$\varphi_s^\mathrm{red}$ of each state $s$ becomes a function of
$\varphi^0$:
\begin{equation}
    \label{eq:PhiZero}
    \varphi_s^\mathrm{red}(\varphi^0)
    =
    \varphi^0
    + \sum_{s' = 1}^S
        \left(
            \Delta \varphi^\mathrm{setup} + \Delta \varphi_{s'}
        \right)
\end{equation}
Inserting Eq.~(\ref{eq:PhiZero}) into Eq.~(\ref{eq:TotalDelayState})
and using Eq.~(\ref{eq:delayCost}) provides us with a periodic
function $D(\varphi^0)$ for the total delays $D$  produced at the
intersection given the start-phase $\varphi^0$. With this function,
the optimal start-phase $\hat \varphi^0$ can be found as $D(\hat
\varphi^0) = \min_\varphi D(\varphi)$. The start-phase $\varphi^0$
can be shifted to the optimal value $\hat \varphi^0$ by an offset to
the input argument of the map, e.g. $s = M(\varphi(t) -
\delta\varphi(t))$.

By adjusting the start-phases, a mutual adaption of the start times
of green-phases at neighbouring intersections is reached. This will
ensure a  minimum delay for all vehicles which have to pass the
intersection. Once the optimum $\hat\varphi^0$ is reached for two
intersections, the switching sequence is repeated periodically until
the traffic situation changes and a new optimum is obtained by the
control concept outlined above. Under certain circumstances, it
results in emergent green waves.

\section{Conclusions}
\label{sec:conclusions}

In this paper we have developed a method to reach coordination among
the traffic lights of an urban street network. This method is based
on a combination of a synchronisation resulting from  local
interactions  and a pure local optimisation:
\begin{enumerate}
\item
At each intersection, the maximum frequency is determined that
allows one to clear the queues on the incoming road sections within
one switching cycle. This frequency is given by Eq.
(\ref{eq:omegaMax}) in dependence of the respective road
utilisation.
\item
The signal controls of all intersections are eventually
adjusted to the minimum of all these frequencies, based on a
coupling of neighbouring signals. A decentralised method of
adjustment is described by Eq.~(\ref{eq:PhaseSync}).
\\
Thereby, all intersections of the road network continuously equalise
the period of their switching cycles. Thus, the switching sequences
and the arrivals rates of the vehicles at all intersections become
periodic functions.
\item
It is always tried to increase this frequency in order to reduce the
waiting times of vehicles in queues. This is reached by introducing
a constant drift, see Eq. (\ref{eq:FrequSync}).
\item
Each cycle is subdivided into different green-phases and setup-times
as illustrated in Fig. 1. Note that, within one switching cycle,
some roads may be served several times.
\item
The optimal green-phases are obtained by minimising the objective
function (\ref{eq:delayCost}). If a periodic sequence of green
lights is used, the optimisation problem becomes a minimisation of a
continuous function with a single parameter $\varphi^0$. In order to
minimise the waiting times of vehicles during red light, the model
presented in Sec.~\ref{sec:traffic} determines the optimal
start-phase $\hat \varphi^0$ of the switching sequence for each
intersection.
\end{enumerate}

Under suitable conditions, our concept allows for the emergence of
green waves. Let's consider the example of a single main road with
several intersections and different average flow rates in both
directions. Then, the vehicles belonging to the highest traffic flow
will always have a green light except for the very first traffic
light where the vehicles queued are bundled. The green wave will
propagate with the largest group of vehicles.

Although our method is quite promising, it has also some
limitations: First, the method for determining the maximum
frequencies $\omega^\mathrm{max}$ needs to be extended for practical
application: In real traffic networks, it is possible that the
traffic arriving at an intersection exceeds capacity. In this case,
it is not possible to remove all waiting vehicles within one cycle.
Within the framework outlined above, the intersection would be just
blocked. However, this situation can be easily avoided by defining a
global minimum of the allowed switching frequencies. If the overload
situation occurs temporarily, such a minimum frequency will still
guarantee a stable functioning of the concept, but it can not avoid
queues that persist for more than one switching cycle. Furthermore,
to dissolve existing queues quickly and to increase the robustness
with respect to stochastic variations in the average flow rates, the
maximum frequencies could be assigned a smaller value than
determined by Eq.~(\ref{eq:omegaMax}).
\\
Second, the frequency of the synchronised system is likely to
decrease with growing system size, so that waiting times can become
quite long. Therefore, synchronisation cannot be the only goal in
signal control. Instead, it would make sense to break up a large
synchronised system into smaller synchronised clusters, which may
change their shape and size in the course of time. It would also be
reasonable to change the order of green-phases sometimes in response
to different traffic patterns (e.g. inbound or outbound traffic
during the morning and afternoon rush hours, respectively).

Nevertheless, it is useful to have a method for a self-organised
coordination of traffic or material flows in a network, based on
local interactions. The next step will be to couple this approach
with a concrete simulation of the dynamics of material flows and
queue formation in the system, as proposed in Ref.
\cite{HelbingLaemmer05}.

In summary, the optimal switching of traffic lights in non-trivial
road networks is one of the most complex material flow control
problems.
However, the  problem in most material flow networks is similar. In
production and manufacturing: road sections in traffic correspond to
buffers in production, travel and delay times to cycle or production
times, junctions to processing units, and different
origin-destination flows in road networks to different product
flows. Thus, our approach is easily transferable to the control of
production systems or other material flow networks. It is
particularly suited for systems of moderate size and load, which
applies to many manufacturing systems.


\section*{Acknowledgements}
\label{sec:acknowledgement}

We thank the participants of the Thematic Institute on Information
and Material Flows in Complex Networks for inspiring discussions, in
particular Dieter Armbruster. D.H. and K.P. kindly acknowledge
partial financial support from the DFG project He 2789/5-1 and the
EU project MMCOMNET. S.L. is grateful for a scholarship by the
Studienstiftung des Deutschen Volkes. H.K is grateful for financial
support by the Alexander von Humboldt foundation.



\end{document}